\documentclass[a4paper,12pt]{article}
\usepackage[]{epsfig}
\usepackage{amsfonts}
\usepackage[latin1]{inputenc}
\usepackage[T1]{fontenc}
\usepackage{amsmath}
\usepackage{amssymb}
\usepackage{color}
\usepackage{graphicx}

\newcommand{\be}{\begin{equation}}
\newcommand{\ee}{\end{equation}}
\newcommand{\ba}{\begin{eqnarray}}
\newcommand{\ea}{\end{eqnarray}}

\newcommand{\PP}{{\cal P}}
\newcommand{\TT}{{\cal T}}
 \newcommand{\bea}{\begin{eqnarray}} \newcommand{\eea}{\end{eqnarray}}

\begin{document}
\title{\bf Path-integral Bosonization of $d=2$ $\PP\TT$ symmetric   models}
\author{C.M. Na\'on and F.A. Schaposnik
\\
{\normalsize  \it Departamento de F\'\i sica, Facultad de Ciencias Exactas,}\\
{\normalsize  \it Universidad Nacional de La Plata}\\
{\normalsize\it Instituto de F\'\i sica La Plata-CONICET, UNLP}\\
{\normalsize\it C.C. 67, 1900 La Plata,
Argentina}}
\date{\today}
\maketitle
\begin{abstract}
We discuss bosonization of non-Hermitian $\PP\TT$ invariant fermion models in  $d=2$
space-time dimensions within the path-integral approach in which the generating
functionals   associated to the fermion and boson models can be related. We first discuss the $\PP\TT$ symmetric Thirring-sine-Gordon connection and then extend the treatment to bosonize the Gross-Neveu model.
\end{abstract}
The possibility that non-Hermitian (NH) quantum Hamiltonians could have real energy eigenvalues was a subject of study since the 1940 decade. Seminal   results were obtained more recently by Bender and collaborators, starting from the work \cite{Bender1998} and followed  by the study of a series of NH Hamiltonian invariant under both parity ($\PP$) and time reversal ($\TT$) transformations leading to real energy eigenvalues and conserved probabilities (see \cite{Bender} and \cite{libro} for a more complete list of references). Among the many applications to quantum field theories, one should mention  the case of NH quantum electrodynamics \cite{BM}-\cite{Milton}, dual quantum field theories \cite{BenderJR}, and many other models.

More recently it was observed  that $\PP \TT$-symmetry in topological insulators guarantees unitarity of states in the bulk but lost at the boundary thus violating the bulk-boundary correspondence of topological phases; in connection with this,  the role of chiral symmetry was also investigated (see \cite{Ghatak} and references therein).
There were also studies of $\PP\TT$ symmetric models in condensed matter physics concerning quantum critical points \cite{Dora}, tight binding models \cite{Luck}, quantum entanglement \cite{nuevo}-\cite{Chen} and quantum dynamics without dissipation \cite{Wang}.

In view of the  recent developments mentioned above, it seems relevant to study boson  and fermion quantum field theories within the path-integral framework which, as recently observed, is particularly useful to study $\PP\TT$  models   for general space-time dimensions \cite{ABS}.
It is the purpose of this work to analyze bosonization of $d=2$ space-time models in which  Lagrangians are modified in order to have, $\PP \TT$ symmetry in the absence of hermiticity. In the path-integral approach this implies to show the  connection between generating functionals  of the boson and fermion models and this will be achieved for different $d=2$ models.

The plan of this work is the following: we first discuss, inspired in ref.\cite{BenderJR}  the connection between the $d=2$, $\PP\TT$ sine-Gordon and Thirring  models  \cite{Coleman}-\cite{Mandelstam},  extending, within the path-integral framework, the bosonization results
presented in \cite{Naon} for the Hermitian models  and then extend the treatment to the case of the  $\PP\TT$ Gross-Neveu   model.

\section*{The $\PP \TT$ sine-Gordon model}
The generating functional    for the  $\PP \TT$ extension of the original Hermitian sine-Gordon model proposed in ref.\cite{BenderJR} reads
\be
Z_{sG}^{\PP \TT} = {\cal N} \int D\phi \exp\left(i\int d^2x L_{sG}^{\PP \TT}\right)
\label{properly}
\ee
with $L_{sG}^{\PP \TT}$ given by
\be
L_{sG}^{\PP \TT} = \frac12 \partial_\mu\phi\partial^\mu \phi -\frac{\alpha_0}{\beta^2}\left(\cos\beta\phi +i\epsilon \sin\beta\phi\right) \label{L}
\ee
Note that we have not included the external source coupled to the scalar field since it does not play any role in the bosonization process.

The addition of the sine term in \eqref{L} breaks the original sine-Gordon model hermiticity but its imaginary coefficient guarantees that $\PP \TT$ symmetry holds.

The interaction term in   $L_{sG}^{\PP \TT}$ can be written as
\be
\frac{\alpha_0}{\beta^2}\left(\cos\beta\phi +i\epsilon \sin\beta\phi\right) =  \frac{\alpha_{0}}{\beta^2}
\frac12\left(
\vphantom{\frac12}(1 + i\epsilon) \exp(i\beta\phi) + (1 - i\epsilon) \exp(-i\beta\phi) \right)
\ee
In order to perform the   path-integral  defining
$Z_{sG}^{\PP \TT}$ by means of a power expansion in $\alpha_0$ we shall then have to consider   vacuum expectations   of the form
\be
  \left\langle \prod_i  \exp\left(i \beta_i   \phi(x_i) \right)  \right\rangle_0,
\label{27}
\ee
 where $\langle ~\rangle_0$ indicates v.e.v.'s for the case of free boson fields $\phi$, and
 \be
 \beta_i  =\pm\beta
 \label{31}
\ee
The result of such v.e.v.'s  involves   scalar propagators
which  have  serious infrared divergences. Indeed, the $d=2$ scalar propagator for a free scalar of mass $\mu$ takes, at short distances the form
\be
\Delta_\mu(x;\mu) = -\frac1{2\pi} \log(c\mu|x|)
\label{nombre}
\ee
where $c = (1/2)\exp\gamma$ with $\gamma$ the Euler constant. One has also to take care of ultraviolet divergencies introducing a cutoff $\Lambda$. Using the Wick product and adopting an appropriate normal ordering (which in the path-integral consists in the addition of appropriate counterterms to kill tadpoles) one can easily see that the v.e.v. in \eqref{27} takes the form
\be
 \left\langle \prod_i  \exp\left(i \beta_{i}   \phi(x_i) \right)  \right\rangle_0 = \exp\left(
 -\frac12\sum_{i,j}\beta_i\beta_j  \Delta_\rho(x_i- x_j;\mu )
 \right)
 \label{finally}
 \ee
where the regularized propagator  $ \Delta_\rho(x;\mu)$ takes the form
 \be
 \Delta_\rho(x;\mu) =
  \Delta(\mu|x|) - \Delta_\rho (\rho|x|)
  \label{propa}
 \ee
 Here $\rho$ is a mass that takes care of divergencies arising when $x \to 0$. Indeed,
 \be
 2\pi\lim_{\mu|x|\to 0 }  \Delta_\rho (|x|) = -\log(\mu c) + \log(\rho c) = -\log(\mu/\rho)
 \label{35}
 \ee
 Now, one has also to take care of ultraviolet divergencies and this is achieved by introducing  a mass $\Lambda$ which will act as a cutoff by introducing a propagator $\tilde \Delta_\rho(x;\mu)$ in the form
 \be
\tilde \Delta_\rho(x;\mu) = \Delta_\rho(x;\mu) -    \Delta_\rho(x;\Lambda)
\ee
so that finally eq.\eqref{finally} takes the form
\bea
&&  \left\langle \prod_i  \exp\left(i \beta_{ i}   \phi(x_i) \right)  \right\rangle_{\!\!0} = \exp\left(
 -\frac12\sum_{i,j}\beta_i\beta_j  \tilde\Delta_\rho(x_i- x_j;\mu )
 \right) \nonumber\\
 = &&   \exp\left(
 -\frac12\sum_{i=j}\beta_i\beta_j  \tilde\Delta_\rho(x_i- x_j;\mu )\right)
 \exp\left(
 -\frac12\sum_{i\ne j}\beta_i\beta_j  \tilde\Delta_\rho(x_i- x_j;\mu ) \right) \nonumber\\
 \label{once}
\eea
where we have separated the $x_j = x_i$ and the $x_j \ne x_i$ factors. Using eq.\eqref{35} the former one takes the form
\bea
\lim_{x_i \to x_j} \exp\left(
-\frac12\sum_{i}\frac{\beta_i^2}{8\pi} \tilde{\Delta} (x_i - x_j)\right)&=&
\left(\frac\mu\rho\right)^{\sum_i \beta_i^2/(8\pi)}  \left(\frac\rho\Lambda
\right)^{-\sum_i\beta_i^2/(8\pi)}\nonumber\\
 &=&  \left(\frac\mu\Lambda\right)^{\left({(\sum_i  \beta_i)^2/(4\pi)}\right)}
\eea
If we now include the $i \ne j$ terms one can prove that the complete  $\mu$-dependent factor becomes
\be
 \langle \exp i{\footnotesize\sum_j}  \beta_j\phi(x_j)  \rangle_0  =
\left(\frac\mu\rho\right)^{\left({(\sum_i  \beta_i)^2/(4\pi)}  + \sum_{i \ne j} (\beta_i\beta_j/4\pi)\tilde\Delta(x_i - x_j)\right)}
\label{result}
\ee

Since bosons $\phi$ are massless one has to take the $\mu \to 0$ limit, this implying that  the v.e.v. vanishes,  except if
\be
\sum_i\beta_i = 0
\label{neutrality}
\ee
Hence, only the $i\ne j$ factor remains. Now, taking $\Lambda|x_i - x_j|  \gg 1$ and $\mu|x_i - x_j| \ll 1$
one has $\Delta(x_i - x_j)$ as given in  eq.\eqref{nombre}.

We can now use result \eqref{result} taking into account eqs.\eqref{27}-\eqref{31} and perform a perturbative expansion in $\alpha_0$ \cite{Naon}. As a consequence of result (\ref{neutrality}) only the terms which are even in $\frac{\alpha_0}{\beta^2}$ contribute to the series, so that the properly normalized generating functional \eqref{properly} takes the form
\bea
Z_{sG}^{\PP \TT} &=&\sum_{k=0}^{\infty}\frac1{k!^2}(\alpha/\beta^2)^{2k}(1 -\epsilon^2)^k
 \nonumber\\
&& \times  \int
\frac{\prod_{i>j}^k
(c^2\rho^2 |x_i - x_j||y_i - y_j|)^{\beta^2/2\pi}
   }
 {\prod_{i,j = 1}^k (  c \rho   |x_i - y_j|)^{\beta^2/2\pi}} \prod_{i=1}^k d^2x^i   d^2y^i
 \label{finale}
 \eea
 where we have defined the renormalized constant
 \be
 \alpha = \frac{\alpha_0}2\left(\frac\rho\Lambda\right)^{\beta^2/4\pi}
 \ee
 \section*{The $\PP \TT$ Thirring model}
Let us start from the generating functional  $Z_{Th}$ for the $\PP \TT$ invariant Thirring model introduced in  ref.\cite{BenderJR},
\be
Z_{Th}^{\PP \TT} = \int D\bar\psi D\psi \exp\left(i\int d^2x L_{Th}^{\PP \TT}\right)\;,
\label{Z}\ee
with the Thirring $\PP \TT$ invariant Lagrangian given by
\be
L_{Th}^{\PP \TT} = \bar\psi (i\not\!\partial - m_1 - m_2\gamma_5)\psi
 +\frac{g^2}2  (\bar\psi\gamma_\mu\psi)(\bar\psi\gamma^\mu \psi)\;,
\label{S}\ee
where we are working in Minkowski space (${\rm diag}\,g_{\mu\nu}=  {(1,-1)}$) with gamma matrices taken as
\be
\gamma_0 = \left(
\begin{array}{cc}
0 & 1\\
1 & 0
\end{array}
\right)\;,
\hspace{0.5cm}
\gamma_1 = \left(
\begin{array}{cc}
0 & -1\\
1 & 0
\end{array}
\right)\;,
\hspace{0.5 cm}
\gamma_5 =  \gamma_0\gamma_1\;, \hspace{0.5cm}  \gamma^\mu\gamma_5 = \varepsilon^{\mu\nu}
\gamma_\mu
\ee
Parity reflection operator $\PP$ and time reversal $\TT$ acting on fermions satisfy
\bea
\PP \psi(x,t)\PP = \gamma^0\psi(-x,t)\;, &&  \PP \bar\psi(x,t)\PP = \bar \psi(-x,t) \gamma^0\\
\TT \psi(x,t)\PP = \gamma^0\psi(x,-t)\;, &&  \TT \bar\psi(x,t)\TT = \bar \psi(x,-t) \gamma^0
\label{PT}
\eea
Note that the $m_2$ term is non-Hermitian because it changes sign under Hermitian conjugation. Now,   being odd both under parity and time reversal transformations  it is $\PP \TT$-symmetric
since $L_{Th}^{\PP \TT} = \PP \TT L_{Th}^{\PP \TT} \PP \TT$.

 Following \cite{Naon}, we now proceed to linearize the quartic fermion interaction by introducing an auxiliary vector field through the identity
\be
\exp\left(i\frac{g^2}2\int d^2x (\bar\psi\gamma_\mu\psi)(\bar\psi\gamma^\mu \psi\right)\!\! =\!\!
\int \!\!DA_\mu \exp\left(\!i\int d^2x  \left(-\frac12 A_\mu A^\mu - g\bar\psi \not\!\!A\psi\right) \right)
\ee
so that the generating  functional \eqref{Z} can be written as
\be
Z_{Th}^{\PP \TT} \!= \int\! D\bar\psi D\psi DA_\mu \! \exp\!\left(\!i\!\int\! d^2x \bar\psi (i\not\!\partial - m_1 - m_2\gamma_5)\psi  -\frac12 A_\mu A^\mu - g\bar\psi \not\!\!A\psi
\right)\!.
\label{Zinter}\ee

Now, in $d=2$ dimensions, the vector field $A_\mu$ can be written in the form
\be
A_\mu = \frac1g \varepsilon_{\mu\nu}\partial^{\nu} \phi + \frac1g \partial _\mu \eta.
\ee
This formula implies that  while $\eta$ is a scalar field,  $\phi$ is a pseudoscalar in view of the Levi-Civita symbol character, as it was originally established within the original (Hermitian) Thirring-sine Gordon bosonization connection \cite{Mandelstam}. Such behavior under $\PP$ is a general feature in any dimensions and follows from the bosonization formula which relates the fermion current to a (pseudo) Kalb-Ramond bosonic dual  \cite{FAS}.

Let us now consider the following change in the path-integral  variables $\bar \psi, \psi \rightarrow \bar \chi, \chi$ and $A_\mu \rightarrow \phi,\eta$ so that fermions change according to
%
 \bea
 \psi(x) = \exp( i\gamma_5\phi(x) +i\eta(x))\chi(x) \nonumber\\
 \bar\psi(x)= \bar\chi(x) \exp(i\gamma_5\phi(x) -i\eta(x))
 \eea
 Now, at the generating  functional level we have to take into account the Fujikawa Jacobian \cite{Fujikawa} $J[\phi]$ which takes the form \cite{Naon}
\be
J[\phi] = \exp\left( {-i} \frac1{2\pi}\int d^2x (\partial_\mu \phi)^2 \right)
\ee
The $\eta$ field completely decouples from fermions so that  it will play no role in the Thirring model bosonization \cite{Naon}. We  shall then ignore it in what follows.
Concerning the $A_\mu \to (\phi,\eta)$ change of variables, it just gives a trivial Jacobian $J_A \propto (1/g^2) \det\nabla^2$ which can be absorbed in the generating functional normalization so that
$Z_{Th}^{\PP \TT}$  reads
 \be
Z_{Th}^{\PP \TT} = \int D\bar\psi D\psi D\phi     \exp\left(i\int d^2x L_{Th}^{\PP \TT}\right)\;,
\label{ZZ}\ee
with $L_{Th}^{\PP \TT}[\bar\chi,\chi,\phi]$
given by
 \be
L_{Th}^{\PP \TT} =   \left( \frac1{2g^2} - \frac1{2\pi}\right) (\partial_\mu\phi)^2   + \bar\chi i\not\!\partial \chi   
- \bar\chi\exp( i\gamma_5\phi) ( m_1 + m_2\gamma_5)\exp(i\gamma_5\phi)\chi
\label{Sdes}
\ee

%

In order
to prove the equivalence between the $\PP \TT$ invariant Thirring and sine-Gordon models    we shall make a perturbative expansion of the former  in the ``mass'' term $\bar\chi M(x)\chi$, where $M(x)$ is the matrix
\be
M(x) \equiv (m_1 +\gamma_5 m_2) \exp(2\gamma_5 \phi)
\ee
At this point, in analogy with the procedure followed for the sine-Gordon model leading to expression (\ref{27}) , choosing an adequate normalization we write $Z_{Th}^{\PP \TT}$ as the following vacuum to vacuum expectation value
\be
Z_{Th}^{\PP \TT}=  \left\langle  \exp\left(-i\int d^2x \bar\chi M(x)\chi \right)  \right\rangle_0,
\label{28}
\ee
 where the v.e.v. above must be computed in a theory of free fermions and free bosons, with propagators
 \be\label{freefermion}
 G_F(x)=-\frac{i}{2\pi} \frac{\gamma^{\mu} x_{\mu}}{x^2},
 \ee
 and
\be\label{freeThirringboson}
\Delta_\mu(x;\mu) = -\lambda^2 \frac1{2\pi} \log(c\mu|x|),
\ee
respectively, with
\be
\lambda^2 = \frac{g^2}{1 - g^2/\pi}.
\ee

In order to get insight into the structure of the perturbative expansion, we shall display terms up to order $M^2$ explicitly. To this end we express the fermionic bispinors as
 \be\chi = \left(
 \begin{array}{c}
 \chi_1\\
 \chi_2
 \end{array}\right)
 \ee
 and $\bar\chi=\chi^{\dagger}\gamma_0$. The first order term reads
 \be
 \bar\chi M \chi  =  (m_1 + m_2)\exp(2i\phi)\bar\chi_1\chi_1+    + (m_1 - m_2)\exp(-2i\phi)\bar\chi_2\chi_2.\label{firstorder}
         \ee

Concerning second order one has, after some work,
 \bea
 \bar\chi M \chi(x) \bar\chi M \chi(y) &=&(m_1^2 - m_2^2) \left(\vphantom{\frac12} A_+(x)A_-(y)\exp(2i(\phi(x) - \phi(y)) \right.\nonumber\\
 &&  + \left.\vphantom{\frac12(m_1^2 - m_2^2) } A_-(x)A_+(y)\exp(-2i(\phi(x) - \phi(y)) \right),
\eea
where we have defined
\be
 A_{\pm} \equiv \bar\chi\frac12 (1 \pm \gamma_5)\chi =\bar\chi_j \chi_j,
\ee
 with $j=1 (2)$ for upper (lower) sign of $A_{\pm}$ (repeated indices do not imply sum in this formula).

 Then, each term in this expansion can be written as the product of a fermionic and a bosonic factor. In particular one can easily see that the expectation value of the first order term (\ref{firstorder}) vanishes. Indeed, in this case, the v.e.v.'s of both fermionic and bosonic factors are zero when an adequate normal ordering that removes Wick contractions of fields at the same point, is adopted. Concerning the evaluation of the bosonic piece, it is completely similar to the one performed for the sine-Gordon model, which led to equation (\ref{neutrality}), the only difference being that in the present case one has $\beta_i= \pm 2$.

In this way one can calculate all orders in $ M(x)$ powers. One finds that the above arguments leading to the vanishing of the first order term can be generalized to all odd orders. Concerning the even ones, all of them contribute to (\ref{28}), and can be readily computed using (\ref{freefermion}) and (\ref{freeThirringboson}). By carefully analyzing the combinatorics of each term, and writing $m_1 = m$ and $m_2 = \epsilon m $ we finally obtain

\be
Z_{Th}^{\PP\TT} = \sum_{k=0}^{\infty}\frac1{k!^2}(m^2(1 -\epsilon^2))^k
 \int
\frac{\prod_{i>j}^k
(c^2\rho^2 |x_i - x_j||y_i - y_j|)^{\beta^2/2\pi}
\prod_{i=1}^k    }
 {\prod_{i,j = 1}^k (  c \rho   |x_i - y_j|)^{\beta^2/2\pi}} d^2x_i   d^2y_i
 \label{44}
\ee

Comparing \eqref{finale} and \eqref{44} one can see that both generating functionals are identical provided the following identification hold,
\bea
&&\frac{\beta^2}{4\pi} = \frac1{1 - g^2/\pi} \\
&&\frac{\alpha}{\beta^2}\cos(\beta\phi) = m\sqrt{1 - \epsilon^2}\bar\psi \psi
\eea
For $\epsilon = 0$ the generating function $Z_{Th}^{\PP \TT}$ just coincides with the generating functional for the Hermitian Thirring model with fermion mass $m$.  Hence as originally found in \cite{BenderJR} one should impose $\epsilon^2 < 1$ in order to have a consistent $\PP \TT$ Thirring massive model with a real fermion mass
$m(1 - \epsilon^2)^{1/2}$.

\section*{The $\PP\TT$ Gross-Neveu model and its bosonization}
We start from the path-integral defining the generating functional   $Z_{GN}$ for the  Hermitian $d=2$ Gross-Neveu  fermion model  %
\be
Z_{GN} = \int D\bar\psi D\psi \exp\left( i\int d^2x L_{GN}[\bar\psi,\psi]\right)\;,
\label{ParFun}
\ee
with $\psi$ a massless fermion and the Lagrangian $L_{GN}$ given by
\be
L_{GN}[\bar\psi,\psi] =  \bar\psi i\!\!\not\!\partial\psi +\frac12 g'^2 (\bar \psi \psi)^2
\ee
Introducing an auxiliary   scalar field $\sigma(x)$, $Z_{GN}$ can be written in the form
\be
Z_{GN}\! = \!\int D\sigma \exp\left(\!-i\int d^2x \frac{\sigma^2}2\right)
\int D\bar\psi D\psi \exp\left(i\int d^2x \left(\bar\psi i\!\!\not\!\partial\psi - g'\bar\psi\psi \sigma \right)\right)
\label{ParFun2}
\ee
or
\be
Z_{GN}  = \int D\sigma \exp\left(\!-i\int d^2x \frac{\sigma^2}2\right)
\det  (i\!\!\not\!\partial - g'  \sigma)
\label{ParFun22}
\ee
with the fermion determinant given by
\be
\det  (i\!\!\not\!\partial - g'  \sigma) = \left\langle \exp\left(-ig'\int d^2x \sigma\bar\psi \psi\rangle\right) \right\rangle_0
\label{Carlos1}
\ee
where $\langle~\rangle_0$ indicates the v.e.v. for the case of free fermions.
Inspired
 in the mass expansion that we performed in the case of the Thirring model  we shall  now proceed, starting from the  r.h.s. of eq.\eqref{Carlos1}  to expand the exponential in powers of $g'$ taking into account that no odd $g'$ powers will contribute after an appropriate tadpole treatment. Moreover, a careful comparison with the Hermitian Thirring model allows to verify that the combinatorial factors of each term coincides with the Thirring case. Then, it becomes apparent that the above fermionic determinant can be expressed as the same  expansion in the Hermitian Thirring coupling constant limit $g \to 0$ and $m \to m(x)\to g'\sigma(x)$.   Thus, one gets
 \be
\det  (i\!\!\not\!\partial - g' \! \sigma) =\!\! \sum_{k=0}^{\infty}\!\! \frac1{k!^2}{g'}^{2k}
 \!\!\int\!\!
\frac{\prod_{i>j}^k
(c^2\rho^2 |x_i - x_j||y_i - y_j|)^{2}
\prod_{i=1}^k    }
 {\prod_{i,j = 1}^k (  c \rho   |x_i - y_j|)^{2}} \sigma(x_i) \sigma(y_i)d^2x_i   d^2y_i
\label{Carlos3}
\ee

We shall now show that the same expansion in eq.\eqref{Carlos3} can be obtained for a sine-Gordon like bosonic model provided the cosine coefficient is $x$-dependent. Indeed, consider the following vacuum functional
\bea
Z_{sG_x}\!\! &=&\!\!\int\!\! D\phi \exp \left(i\int d^2x \frac{1}{2}\partial_\mu\phi\,\partial^\mu\phi\right)
 \exp\left(i\frac{\alpha_0(x)}{\beta^2}\cos(\beta\phi)\right)\nonumber\\
 &=& \left\langle \exp\left(i\int d^2x \frac{\alpha_0(x)}{\beta^2}\cos(\beta\phi)\right)
 \right\rangle_0
 \label{arriba}
 \eea
where now $\langle~\rangle_0$ is the v.e.v. for free bosons.

To see this connection we make  an expansion in powers of
$\alpha_0(x)$ in the r.h.s. of \eqref{arriba}.   Again, odd powers   vanish so that, for the v.e.v. \eqref{arriba} we get an expression similar to the one corresponding to the usual, Hermitian sine-Gordon model, but with the coefficient $\alpha$ promoted to be a function of $x$:

\bea
Z_{sG_x} &=&\sum_{k=0}^{\infty}\frac1{k!^2}(1/\beta^2)^{2k}
 \times \nonumber\\
&&  \int
\frac{\prod_{i>j}^k
(c^2\rho^2 |x_i - x_j||y_i - y_j|)^{\beta^2/2\pi}
   }
 {\prod_{i,j = 1}^k (  c \rho   |x_i - y_j|)^{\beta^2/2\pi}} \prod_{i=1}^k \alpha(x_i)\alpha(y_i) d^2x_i   d^2y_i
 \label{finale-x}
 \eea
 where we have defined the renormalized function
 \be
 \alpha(x) = \frac{\alpha_0(x)}2\left(\frac\rho\Lambda\right)^{\beta^2/4\pi}.
 \ee

Comparing this result with that in \eqref{Carlos3} for the fermion model, we see that both expansions coincide  provided the following identities hold:
 \bea
 \frac{\beta^2}{2\pi} &=& 2 \nonumber\\
 g \sigma(x) &=&\frac{\alpha(x)}{\rho}.
 \label{id}
 \eea

We have then that the GN fermion model can be bosonized becoming the sG$_x$ boson model since their generating functionals have been identified
\be
Z_{GS_x}[\bar\psi, \psi,\sigma]= Z_{GN}[\phi]
\ee
Hence,  together with  identities \eqref{id} the bosonization rule for the interaction  terms reads
\be
g\sigma(x)\bar\psi\psi =
 \frac{\alpha(x)}{4\pi}  \cos(2\sqrt\pi\ \phi)
 \ee

Up to now we have considered the path-integral bosonization of the usual, Hermitian Gross-Neveu model, which was not previously discussed in the literature. Concerning the connection for the case of $\PP\TT$ symmetric models,   from what we learnt for the case of $\PP\TT$ sine-Gordon-Thirring equivalence, it is evident that for  the case of the sine-Gordon model with an $\alpha(x)$  dependent coefficient, in order to define a $\PP\TT$ symmetric model
one should extend \eqref{arriba} as
\be
Z_{sG_x}^{\PP\TT} = \!\!\int D\phi \exp \left(\!i\int d^2x \left(\frac{1}{2}\partial_\mu\phi\,\partial^\mu\phi
 + \frac{\alpha(x)}{4\pi}(\cos(\beta\phi(x))  + i \epsilon \sin(\beta\phi(x))\right) \right)
 \ee
 with $\beta = 2\sqrt\pi$.  {Then, it is easy to show   that the bosonic partition function
 $Z_{sG_x}^{\PP\TT}$
is equivalent to a fermionic $\PP\TT$ symmetric Gross-Neveu model that reads}
\be
Z_{GN}^{\PP\TT}\!\! = \!\!\int\!\! D\sigma \exp\left(\!\!-i\int\!\! d^2x \frac{\sigma^2}2\!\right)\!\!
\int\!\! D\bar\psi D\psi \exp\left(\!\!  i\int\!\!\!d^2x \!\!\left(\bar\psi i\!\!\not\!\partial\psi - g'\bar\psi\psi \sigma\sqrt{1-\epsilon^2}\, \right)\!\right)\!.
\label{ParFun2pt}
\ee

This result completes our discussion on the equivalence between non Hermitian fermionic and bosonic 2D models with $\PP\TT$ symmetry, through path-integral methods.

In summary, we have presented a path-integal approach to bosonization of $\PP\TT$ models complementary to the one originally discussed in \cite{BenderJR}. Concerning the sine-Gordon-Thirring connection we have explicitly shown the identity between the resulting generating  functionals $Z_{Sg}^{\PP\TT}$ and $Z_{Th}^{\PP\TT}$, the path-integral analog of the well-honored Coleman perturbation series results for the Hermitian models \cite{Coleman}. Using the same approach we have also discussed bosonization in the case of the Gross-Neveu model finding the corresponding generating functionals of the dual boson and fermion $\PP\TT$ symmetric models.

In view of the recent applications in condensed matter physics discussed in the introduction we expect to extend our path integral bosonization approach  to  $d=3$ dimensional fermionic  models along the lines of the Hermitian case described in \cite{FAS},  in view to describe $\PP\TT$ planar models.

\section*{Acknowledgments}
We would like to thank Eduardo Fradkin for discussions and Daniel Cabra and An\'\i bal Iucci for   helpful suggestions.  F.A.S  research was supported by  PIP 02229 and UNLP X910 grants. C.M.N. research was supported by UNLP X917 grant.

\end{document}